\documentclass[aps,prb,superscriptaddress,twocolumn,floatfix]{revtex4-1}
\usepackage{graphicx}
\usepackage{amssymb}
\usepackage{amsfonts}
\usepackage{amsmath}
\usepackage{epstopdf}

\begin{document}

\title{High-pressure study of the basal-plane anisotropy of the upper critical field of the topological superconductor Sr$_x$Bi$_2$Se$_3$}

\author{A. M. Nikitin} \email{a.niktin@uva.nl}\affiliation{Van der Waals - Zeeman Institute, University of Amsterdam, Science Park 904, 1098 XH Amsterdam, The Netherlands}
\author{Y. Pan} \affiliation{Van der Waals - Zeeman Institute, University of Amsterdam, Science Park 904, 1098 XH Amsterdam, The Netherlands}
\author{Y. K. Huang} \affiliation{Van der Waals - Zeeman Institute, University of Amsterdam, Science Park 904, 1098 XH Amsterdam, The Netherlands}
\author{T. Naka} \affiliation{National Institute for Materials Science, Sengen 1-2-1, Tsukuba, Ibaraki 305-0047, Japan}
\author{A. de Visser} \email{a.devisser@uva.nl} \affiliation{Van der Waals - Zeeman Institute, University of Amsterdam, Science Park 904, 1098 XH Amsterdam, The Netherlands}

\date{\today}

\begin{abstract}
We report a high-pressure transport study of the upper-critical field, $B_{c2}(T)$, of the topological superconductor Sr$_{0.15}$Bi$_2$Se$_3$ ($T_c = 3.0$~K). $B_{c2}(T)$ was measured for magnetic fields directed along two orthogonal directions, $a$ and $a^*$, in the trigonal basal plane. While superconductivity is rapidly suppressed at the critical pressure $p_c \sim 3.5$~GPa, the pronounced two-fold basal-plane anisotropy $B_{c2}^a/B_{c2}^{a^*} = 3.2$ at $T=0.3$~K, recently reported at ambient pressure (Pan \textit{et al}., Ref.~\onlinecite{Pan2016}), is reinforced and attains a value of $\sim 5$ at the highest pressure (2.2 GPa). The data reveal that the unconventional superconducting state with broken rotational symmetry is robust under pressure.
\end{abstract}

\pacs{74.70.Dd, 74.25.Op, 74.62.Fj}

\maketitle

\section{INTRODUCTION}

The tetradymite Bi$_2$Se$_3$ is one of the prototypical materials that played an instrumental role in developing the field of 3-dimensional topological insulators~\cite{Hasan&Kane2010,Qi&Zhang2010}. Electronic structure calculations~\cite{Zhang2009} predicted Bi$_2$Se$_3$ has a non-trivial topology of the electron bands due to large spin-orbit coupling. The bulk of the crystal is insulating and at the surface gapless states exist that are protected by symmetry. The topological surface states are characterized by a helical Dirac-type energy dispersion with the spin locked to the momentum. The topological properties have experimentally been confirmed by Angle Resolved Photo Emission Spectroscopy (ARPES)~\cite{Xia2009,Hsieh2009}. Most interestingly, the topological insulator Bi$_2$Se$_3$ can relatively easily be transformed into a superconductor with $T_c \sim 3$~K by doping with Cu~\cite{Hor2010}, Sr~\cite{Liu2015}, Nb~\cite{Qiu2015} or Tl~\cite{Wang2016}. Making use of the direct analogy of the Bogoliubov-de Gennes Hamiltonian for the quasiparticles of a superconductor and the Bloch Hamiltonian for the insulator it is predicted that these doped systems are topological superconductors~\cite{Hasan&Kane2010,Qi&Zhang2010}. Taking into account the Fermi surface topology in the normal state this can give rise to an odd-parity Cooper pairing symmetry and a fully gapped superconducting state~\cite{Sato2010,Fu&Berg2010}. Among the Bi$_2$Se$_3$-based superconductors, Cu$_x$Bi$_2$Se$_3$ has been studied most intensively~\cite{Hor2010,Kriener2011a,Sasaki2011,Sasaki&Mizushima2015}. A topological superconducting state was concluded based on a two-orbital model for centro-symmetric superconductors exhibiting strong spin-orbit coupling. The possible superconducting order parameters were evaluated by symmetry-group classification ($D_{3d}$ point group, $R \overline{3}m$ space group) and an interorbital spin-triplet state ($\Delta _2$-pairing) was put forward as order parameter~\cite{Fu&Berg2010,Sasaki2011}.

An exciting development in the field of Bi$_2$Se$_3$-based superconductors is the experimental observation of rotational symmetry breaking: a magnetic field applied in the trigonal basal-plane spontaneously lowers the symmetry to two-fold~\cite{Matano2016,Pan2016}. In Cu$_x$Bi$_2$Se$_3$ this was demonstrated for the spin-system by the angular variation of the Knight shift measured by Nuclear Magnetic Resonance (NMR)~\cite{Matano2016}, while specific heat measurements show it is a thermodynamic bulk feature~\cite{Yonezawa2016}. In Sr$_x$Bi$_2$Se$_3$ rotational symmetry breaking was detected by the angular variation of the upper critical field, $B_{c2}(\theta)$, probed by magnetotransport~\cite{Pan2016}, and in Nb$_x$Bi$_2$Se$_3$ by torque magnetometry that senses the magnetization of the vortex lattice~\cite{Asaba2016}. The  rotational symmetry breaking appears to be ubiquitous in Bi$_2$Se$_3$-based superconductors and puts important constraints on the possible order parameters. According to recent models~\cite{Nagai2012,Fu2014,Venderbos2015} it restricts the order parameter to an odd-parity two-dimensional representation, $E_u$, with $\Delta _4$-pairing, which involves a nematic director that breaks the rotational symmetry when pinned to the crystal lattice. This unconventional superconducting state is referred to as nematic superconductivity.

Here we report a high-pressure magnetotransport study on single-crystalline Sr$_{0.15}$Bi$_2$Se$_3$, conducted to investigate the robustness of the rotational symmetry breaking to hydrostatic pressure. The upper-critical field, $B_{c2}(T)$, was measured for magnetic fields directed along two orthogonal directions, $a$ and $a^*$, in the trigonal basal plane. While superconductivity is rapidly depressed with pressure, the pronounced two-fold basal-plane anisotropy $B_{c2}^a/B_{c2}^{a^*} =3.2$ observed at $T=0.3$~K at ambient pressure~\cite{Pan2016}, is reinforced and attains a value of $\sim 5$ at the highest pressure (2.2~GPa). The rapid depression of $T_c$ points to a critical pressure for the suppression of superconductivity $p_c \sim 3.5$~GPa. Recently, a similar rapid decrease of $T_c$ has been reported in a systematic study of the electronic and structural properties of single crystals of Sr$_{0.065}$Bi$_2$Se$_3$~\cite{Zhou2016}. In the pressure range covering $p_c$ the $R \overline{3}m$ space group is preserved. By increasing the pressure further two structural phase transitions are observed, namely at 6~GPa to the $C2/m$ phase and at 25~GPa to the I4/mmm phase. We remark that at pressures above 6~GPa superconductivity reemerges with a maximum $T_c \sim 8.3$~K~\cite{Zhou2016}. This is analogous to the emergence of superconductivity under pressure in undoped Bi$_2$Se$_3$~\cite{Kirshenbaum2013}.

\section{EXPERIMENTAL}

For the preparation of Sr$_{x}$Bi$_{2}$Se$_{3}$ single crystals with a nominal value $x=0.15$, high-purity elements were melted at 850 $^{\circ}$C in sealed evacuated quartz tubes. Crystals were formed by slowly cooling to 650~$^{\circ}$C at a rate of 3~$^{\circ}$C/hour. Powder X-ray diffraction confirmed the R$\bar{3}$m space group. The single-crystalline nature of the crystals was checked by Laue back-reflection. Flat samples with typical dimensions $0.3 \times 2.5 \times 3$~mm$^{3}$ were cut from the bulk crystal with a scalpel blade. The plane of the samples contains the $a$- and $a^*$-axis. The $a$-direction was taken along the long direction of the sample. The characterization of the single-crystalline batch with $x=0.15$ by means of transport and ac-susceptibility measurements is presented in Ref.~\onlinecite{Pan2016}. The superconducting shielding fraction amounts to 80~\%.

High-pressure magnetotransport measurements were carried out with help of a hybrid clamp cell made of
NiCrAl and CuBe alloys. Two crystals were mounted on a plug that was placed in a Teflon cylinder with Daphne oil 7373 as hydrostatic pressure transmitting medium. For resistance measurements in a 4-point configuration thin gold wires were attached to the flat sides of the crystals by silver paste. The magnetic field was applied in the $aa^*$-plane of the sample with configurations $B \parallel a \parallel I$ (crystal 1) and $B \parallel a^* \perp I$ (crystal 2). The pressure cell was attached to the cold plate of a $^3$He refrigerator (Heliox, Oxford Instruments) equipped with a superconducting solenoid ($B_{max} = 14$~T). The effective pressure was determined in previous experiments~\cite{Bay2012a,Bay2012b} and the maximum pressure reached is 2.2~GPa.  Low temperature, $T = 0.24 - 10$~K, resistance measurements were performed using a low-frequency lock-in technique with low excitation currents ($I \leq 100$~$\mu$A).

\section{RESULTS}

\begin{figure}
\includegraphics[width=8cm]{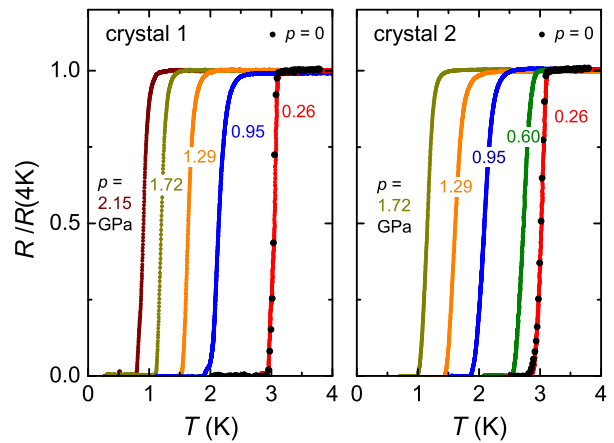}
\caption{Resistance of Sr$_{0.15}$Bi$_{2}$Se$_{3}$ as a function of temperature around ${T_{c}}$ for crystal 1 (left panel) and crystal 2 (right panel) at pressures up to 2.15 GPa as indicated. The resistance is normalized to $R(4$K) at ambient pressure. The black solid circles represent $R(T)$ at ambient pressure.}
\end{figure}

\begin{figure}
\includegraphics[width=8cm]{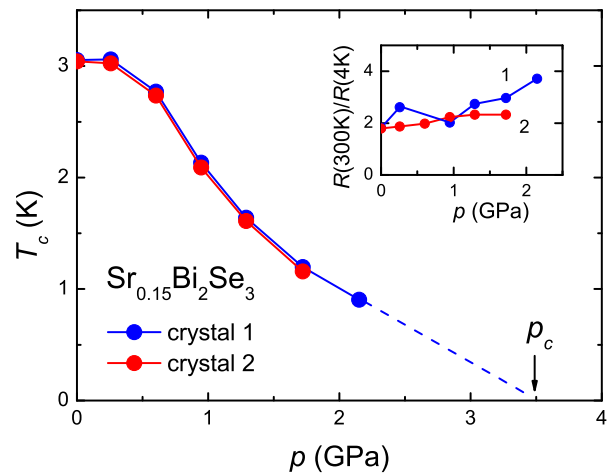}
\caption{Superconducting transition temperature ${T_{c}}$ as a function of pressure for crystal 1 (blue symbols) and crystal 2 (red symbols). The dashed line represents a linear extrapolation of ${T_{c}(p)}$ with ${p_{c} \sim}$ 3.5 GPa. Inset: Normalized resistance $R$(300K)/$R$(4K) as a function of pressure for both crystals.}
\end{figure}

Before mounting the crystals in the pressure cell the temperature variation of the resistivity was measured at ambient pressure. The resistivity shows a metallic behavior and levels off below $\sim 10$~K (see Ref.~\onlinecite{Pan2016}). For both crystals $T_c = 3.05 \pm 0.10$~K as identified by the midpoints of the transitions in $R(T)$ (see black solid circles in Fig.~1). Under pressure the resistivity remains metallic and the resistance ratio $R(300$K$)/R(4$K$)$ increases by $\sim 100$~\% for crystal~1 and $\sim 30$~\% for crystal~2, as shown in the inset of Fig.~2. The larger increase for crystal~1 is mainly due to the decrease of $R(4$K), while $R(4$K) is close to constant for crystal~2. $R(T)$ around the superconducting transition under pressure is shown in Fig.~1 (solid lines). Superconductivity progressively shifts to lower temperatures. In Fig.~2 we trace $T_c (p)$ of both crystals. $T_c$ is smoothly depressed to a value of 0.90 K at 2.15~GPa. The dashed line in Fig.~2 represents a linear extrapolation of $T_c (p)$ and indicates the critical pressure $p_c$ for the suppression of superconductivity is $\sim 3.5$~GPa. We remark that the value $p_c = 1.1$~GPa reported in Ref.~\onlinecite{Zhou2016} most likely underestimates $p_c$ as it is based on the extrapolation of $T_c (p)$ from temperatures above 2 K only. In the case of Cu$_{x}$Bi$_{2}$Se$_{3}$ the critical pressure is estimated to be a factor 2 larger than in the Sr doped case, $p_c \sim 6.3$~GPa~\cite{Bay2012b}.

\begin{figure}
\includegraphics[width=8cm]{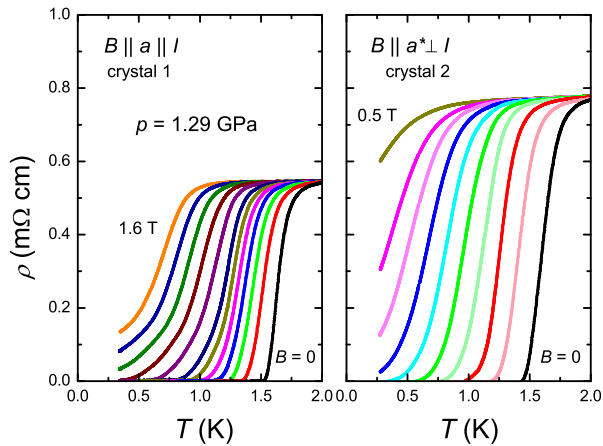}
\caption{Left panel: Resistivity of Sr$_{0.15}$Bi$_{2}$Se$_{3}$ as a function of temperature at $p = 1.29$~GPa for $B \parallel a \parallel I$ (crystal 1) measured in fixed magnetic fields. Curves from right to left: from 0~T to 0.6~T with 0.1~T steps and from 0.8~T to 1.6~T with 0.2~T steps. Right panel: Data for $B \parallel a^* \perp I$ (crystal 2) measured in fixed magnetic fields. Curves from right to left: from 0~T to 0.4~T with 0.05~T steps and in 0.5~T.}
\end{figure}

The resistance as a function of temperature in fixed magnetic fields applied along the $a$- and $a^*$-axis was measured to determine the upper-critical field, $B_{c2}$. Typical data at $p = 1.29$~GPa are presented in Fig.~3. The superconducting transition becomes broader in magnetic field. In order to systematically determine $T_c (B)$ (or $B_{c2}(T)$) we collected the midpoints of the superconducting transitions in $R(T)$. We remark that other definitions of $T_c$, such as a 10\% or 90\% drop of the resistance with respect to the normal state value, will affect the absolute value of $B_{c2}$, but not our central conclusion that the anisotropy of $B_{c2}$ is robust under pressure.

The main results are presented in Fig.~4, where we have plotted $B_{c2}(T)$  at different pressures for $B \parallel a$ (crystal 1) and $B \parallel a^*$ (crystal 2). Note the difference of a factor 2 in the units along the vertical axis between the left and right panel. The most striking feature is the strong depression of $B_{c2}^{a}(T)$ and $B_{c2}^{a^*}(T)$ with pressure. A second remarkable feature is that the strong anisotropy $B_{c2}^a (T) \gg B_{c2}^{a^*} (T)$ persists under pressure.

\begin{figure}
\includegraphics[width=8cm]{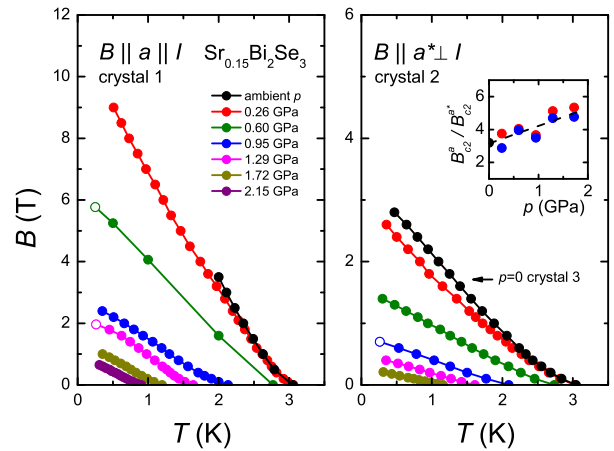}
\caption{Temperature variation of the upper-critical field, $B_{c2}$, of Sr$_{0.15}$Bi$_{2}$Se$_{3}$  with configuration $B \parallel a \parallel I$ (left panel) and $B \parallel a^* \perp I$ (right panel) at pressures (from top to bottom) of 0, 0.26, 0.60, 0.95, 1.29, 1.72 and 2.15 GPa. Open circles yield $B_{c2}$ taken from field sweeps at fixed temperature. The data at $p=0$ (crystal 3) in the right panel are taken from Ref.~\onlinecite{Pan2016}. Inset: Pressure variation of the basal-plane anisotropy $B_{c2}^a/B_{c2}^{a^*}$ at $T = 0.5$~K (solid red symbols) and at $T/T_c = 0.28$ (solid blue symbols). The ambient pressure point is taken from Ref.~\onlinecite{Pan2016}. The dashed line is a guide to the eye.}
\end{figure}

\section{ANALYSIS AND DISCUSSION}

The robustness of the large basal-plane anisotropy can be quantified by tracing the ratio $B_{c2}^a / B_{c2}^{a^*}$ at the low temperature of 0.5~K as a function of pressure (see inset in Fig.~4). This shows the anisotropy ratio is reinforced under pressure and increases by a factor close to 2 at the highest pressure (2.2 GPa). Since $B_{c2}(T)$ does not level off when $T \rightarrow 0$, we have also traced the anisotropy ratio at the reduced temperature $T/T_c = 0.28$ in the inset. The increase is the same. As discussed in Pan \textit{et al}.~(Ref.~\onlinecite{Pan2016}) the large two-fold basal-plane anisotropy cannot be explained by the Ginzburg-Landau anisotropic effective mass model or by the effect of flux flow on $B_{c2}$ due to the Lorentz force for $B \perp I$. Instead it provides solid evidence for unconventional superconductivity. This was put on firm footing by calculations of the upper-critical field in the framework of the Ginzburg-Landau theory for superconductors with $D_{3d}$ crystal symmetry~\cite{Krotkov2002,Venderbos2016}. While for one-dimensional representations $B_{c2}$ is isotropic in the basal plane, two-dimensional representations, like $E_u$, can give rise to a six-fold anisotropy. The further reduction of $B_{c2}$ to two-fold as observed in experiments~\cite{Matano2016,Yonezawa2016,Pan2016,Asaba2016} is then naturally explained by the lifting of the degeneracy of the two components. The origin of the lifting of the degeneracy has not been established yet, and is the subject of future research. Possibly, uniaxial strain or a preferred ordering of the dopant atoms in the Van der Waals gaps between the quintuple layers of the Bi$_2$Se$_3$ crystal, lowers the symmetry.

We remark that both $B_{c2}^a$ and $B_{c2}^{a^*}$ show an unusual temperature variation, notably at low pressures a pronounced upward curvature below $T_c$ is followed by a quasi-linear behavior. In an attempt to model $B_{c2}(T)$ we present the data in Fig.~5 in a reduced plot $b^*(t)$, with $b^* = - (B_{c2}(T)/T_{c})/(dB_{c2}/dT)|_{T_c}$ and $t = T/T_c$ the reduced temperature. In calculating the slope $(dB_{c2}/dT)|_{T_c}$ we have neglected the low-field curvature which resulted in a lower $T_c$. Therefore the reduced data show a tail that extends to $t > 1$. Apart from the low-field curvature that weakens under pressure, the data overall collapse onto a single function $b^*(t)$, like reported, for instance, for Cu$_x$Bi$_2$Se$_3$~\cite{Bay2012a} and the half-Heusler superconductor YPtBi~\cite{Bay2012b}. This shows the functional dependence does not change under pressure. For comparison we have plotted in Fig.~5 the $B_{c2}$-curve for a standard weak-coupling BCS spin-singlet superconductor with orbital limit $B_{c2}^{orb}(0)= - 0.72 \times T_c (dB_{c2}/dT)|_{T_c}$, i.e. the Werthamer-Helfand-Hohenberg (WHH) model curve~\cite{Werthamer1966}. Clearly, the data deviate strongly from the model curve. This calls for detailed theoretical work to model $B_{c2}(T)$ in the case of $\Delta _4$-pairing in the $E_u$ representation. Finally, we remark that the phenomenological Ginzburg-Landau model curve, $B_{c2}(T) = B_{c2}(0)[(1-t^2)/(1+t^2)]$, that was used in Ref.~\onlinecite{Shruti2015} to provide an estimate of $B_{c2}(0)$, yields a description of the collapsed data over fairly large temperature range as shown in Fig.~5, where we used $b^*(0)=0.84$.

\begin{figure}
\includegraphics[width=8cm]{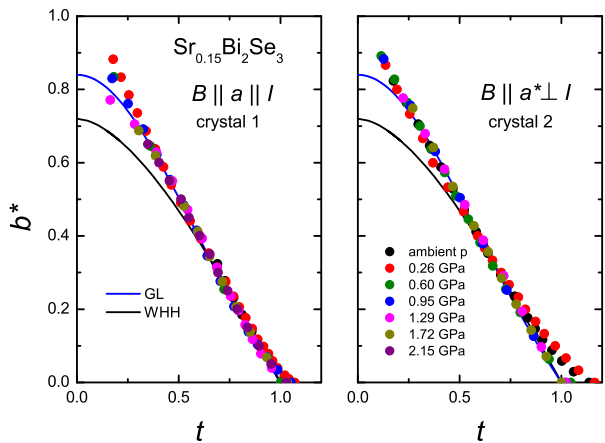}
\caption{Reduced plot $b^*(t)$ of the upper-critical field, with $b^* = - (B_{c2}(T)/T_{c})/(dB_{c2}/dT)|_{T_c}$  and $t = T/T_{c}$ at pressures of 0, 0.26, 0.60, 0.95, 1.29, 1.72 and 2.15 GPa for $B \parallel a^* \perp I$ (left panel) and $B \parallel a^* \perp I$ (right panel). The black solid line represents a comparison with the WHH model for a weak-coupling spin-singlet superconductor and the blue solid line represents the generalized Ginzburg-Landau model (see text). In the determination of the slope $(dB_{c2}/dT)|_{T_c}$ and $T_c$ the low-field curvature of $B_{c2}$ is neglected.}
\end{figure}

\section{Conclusions}

We have carried out a high-pressure study of the topological superconductor Sr$_{0.15}$Bi$_2$Se$_3$ in order to investigate the unusual basal-plane anisotropy of the upper critical field, recently detected at ambient pressure~\cite{Pan2016}. The superconducting transition temperature is rapidly depressed with a critical pressure of $\sim 3.5$~GPa. $B_{c2}(T)$ has been determined for the $B$-field applied along two orthogonal directions, $a$ and $a^*$, in the basal plane. The pronounced two-fold basal-plane anisotropy $B_{c2}^a/B_{c2}^{a^*} =3.2$ at $T=0.3$~K is robust under pressure and attains a value of $\sim 5$ at the highest pressure (2.2 GPa). The two-fold anisotropy of $B_{c2}(T)$ provides solid evidence for rotational symmetry breaking in the $D_{3d}$ crystal structure. This puts stringent conditions on the superconducting order parameter, namely it should belong to a two-dimensional representation ($E_u$). Rotational symmetry breaking seems to be ubiquitous in the family of doped Bi$_2$Se$_3$-based superconductors. This offers an exciting new opportunity to study unconventional superconductivity with a two-component order parameter.

\noindent
\textbf{Acknowledgements}
The assistance of G. K. Araizi in taking part of the data is gratefully acknowledged. This work was part of the research program on Topological Insulators funded by FOM (Dutch Foundation for Fundamental Research of Matter).

\bibliography{Refs_SrBi2Se3}

\bibliographystyle{apsrev4-1}

\end{document}